\documentclass[a4paper,10pt]{IEEEtran}
\usepackage[T1]{fontenc}
\pdfoutput=1
\IEEEoverridecommandlockouts
\makeatletter
\def\endthebibliography{%
	\def\@noitemerr{\@latex@warning{Empty `thebibliography' environment}}%
	\endlist
}
\makeatother

\IEEEoverridecommandlockouts
\usepackage{setspace}
\usepackage{multirow}
\usepackage{lipsum}
\usepackage{amsfonts}
\usepackage{array}
\usepackage{amsmath,cases,amssymb}
\usepackage{amsmath}
\usepackage{amsthm}
\usepackage{graphicx}
\usepackage{cite}
\usepackage{mathtools}
\usepackage{subfigure}
\usepackage{epsfig}
\usepackage{url}
\usepackage{algorithm}
\usepackage{algorithmic}
\usepackage{epstopdf}
\usepackage{color}
\usepackage{makecell}

\usepackage{lipsum}

\DeclareGraphicsExtensions{.eps,.pdf}

\newcommand{\bPsi}{\boldsymbol{\Psi}}

\newcommand{\bepsilon}{\boldsymbol{\xi}}

\newcommand{\bw}{\mathbf{w}}

\newcommand{\bC}{\mathbf{C}}

\newcommand{\bH}{\mathbf{H}}

\newcommand{\bI}{\mathbf{I}}

\newcommand{\bv}{\mathbf{v}}

\newcommand{\bu}{\mathbf{u}}

\newcommand{\bP}{\mathbf{P}}

\newcommand{\bo}{\mathbf{o}}
\newcommand{\bb}{\mathbf{b}}
\newcommand{\ba}{\mathbf{a}}
\newcommand{\bs}{\mathbf{s}}
\newcommand{\bK}{\mathbf{K}}

\newcommand{\bx}{\mathbf{x}}

\newcommand{\bmu}{\pmb{\mu}}

\newcommand{\Qcal}{\mathcal{Q}}
\newcommand{\Kcal}{\mathcal{K}}

\newcommand{\Mcal}{\mathcal{M}}

\newcommand{\Ncal}{\mathcal{N}}

\newcommand{\Pcal}{\mathcal{P}}

\newcommand{\PA}{\mathtt{PA}}
\newcommand{\SA}{\mathtt{SA}}

\makeatletter
\newcommand{\doublewidetilde}[1]{{%
		\mathpalette\double@widetilde{#1}}}
\newcommand{\double@widetilde}[2]{%
	\sbox\z@{$\m@th#1\widetilde{#2}$}%
	\ht\z@=.5\ht\z@
	\widetilde{\box\z@}}
\makeatother

\newtheorem{lemma}{Lemma}

\newtheorem{remark}{Remark}

\begin{document}
	
	\title{Scheduling Policy for Value-of-Information (VoI)
in Trajectory Estimation for Digital Twins}%

\author{Van-Phuc Bui, Shashi Raj Pandey, \textit{Member, IEEE}, Federico Chiariotti, \textit{Member, IEEE}, and Petar Popovski, \textit{Fellow, IEEE}\thanks{V. Bui, S.R. Pandey, F. Chiariotti, and P. Popovski (emails: \{vpb, srp, fchi, petarp\}@es.aau.dk) are all with the Department of Electronic Systems, Aalborg University, Denmark. F. Chiariotti is also with the Department of Information Engineering, University of Padova, Italy. This work was supported by the Villum Investigator Grant ``WATER'' from the Velux Foundation, Denmark.}}

\maketitle
	\vspace*{-0.9cm}
\begin{abstract}
This paper presents an approach to schedule observations from different sensors in an environment to ensure their timely delivery and build a digital twin (DT) model of the system dynamics. At the cloud platform, DT models estimate and predict the system's state, then compute the optimal scheduling policy and resource allocation strategy to be executed in the physical world.
However, given limited network resources, partial state vector information, and measurement errors at the distributed sensing agents, the acquisition of data (i.e., observations) for efficient state estimation of system dynamics is a non-trivial problem. We propose a Value of Information (VoI)-based algorithm that provides a polynomial-time solution for selecting the most informative subset of sensing agents to improve confidence in the state estimation of DT models. Numerical results confirm that the proposed method outperforms other benchmarks, reducing the communication overhead by half while maintaining the required estimation accuracy.
\end{abstract}

\section{Introduction}\label{sec:intro}


The Industry 4.0 smart manufacturing paradigm requires large quantities of real-time data generated from a myriad of wireless sensors\cite{tang2015tracking}. Modern Industrial Internet of Things (IIoT) networks deploy a large number of nodes, which either exchange their noisy (and possibly processed) observations of a random process, act as relays and data fusion nodes by aggregating observations from sensors, or execute control actions.  Digital twin (DT) models transform these large volumes of data into predictive models, which can simulate the consequences of potential control strategies and help system operators decide the best course of action~\cite{9899718}. 

\begin{figure}[t!]
	\centering
	\includegraphics[width=0.45\textwidth]{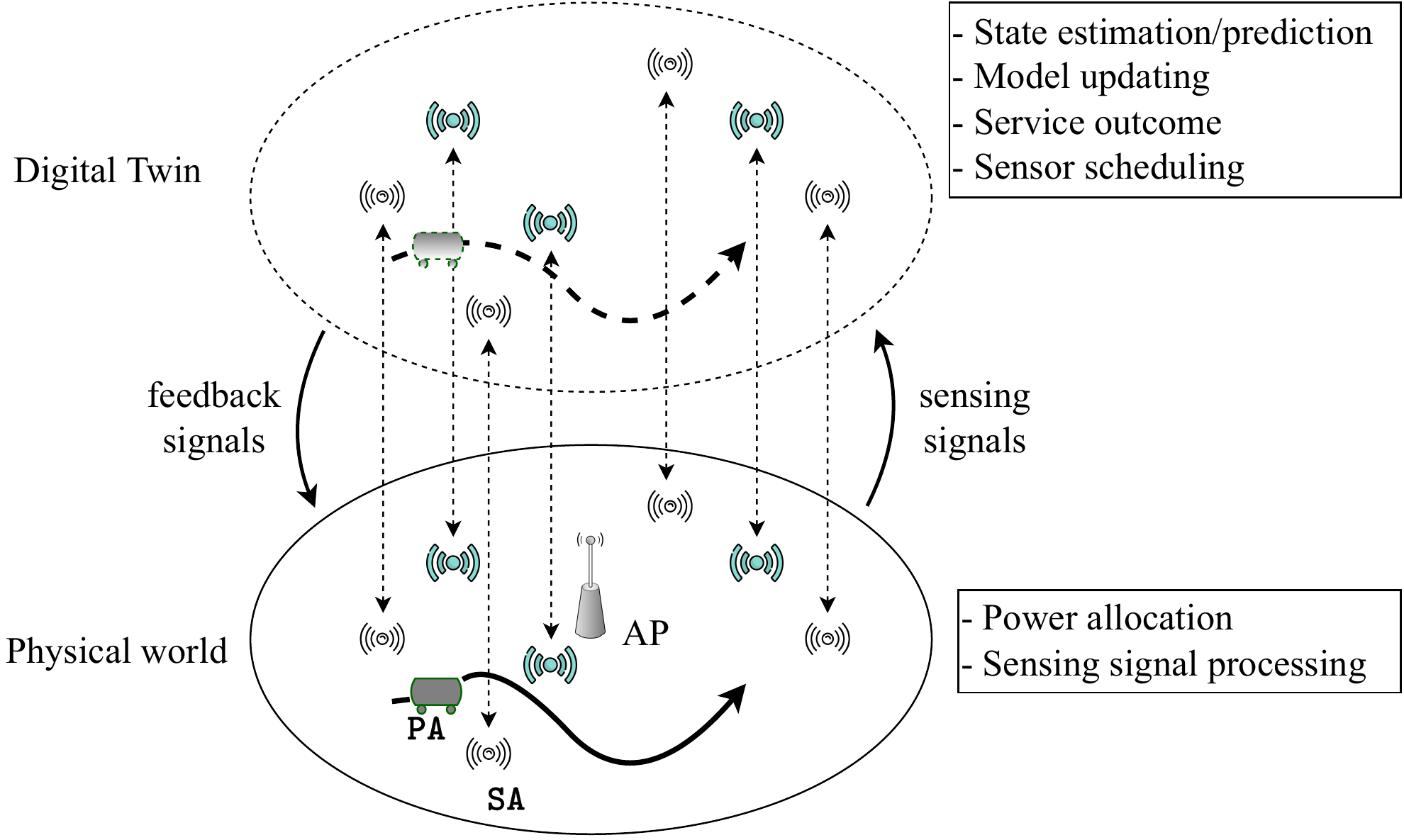}
	\vspace*{-0.3cm}
	\caption{An example of a DT system.}
	\label{fig_system_model}
	\vspace*{-0.65cm}
\end{figure}

Considering the tight interaction between the communication and control systems, it is difficult to design them jointly in order to maintain the predictive performance of the DT model while prolonging the network lifetime. Take into account a system of process dynamics as in Fig.~\ref{fig_system_model} where a DT model is built to estimate the dynamic process of a \emph{primary agent} ($\PA$) interacting with the environment. The  \emph{sensing agents}  ($\SA$s) convey their noisy and partial observations via wireless transmission to an \emph{access point} (AP) directly connected to a Cloud platform \cite{srivastava2018review} for building DT models \cite{li2020digital}. 
After processing at the virtual world using the Cloud platform, i.e., after the model is updated, the next system state is predicted and the optimal policy is computed, commands are sent to be executed in the physical world.
	A DT may model a wireless sensor network (WSN) in which $\SA$s observe various features in the $\PA$'s state, including position, velocity, and acceleration. These types of measurements are used in a wide range of indoor and outdoor applications such as localization and navigation \cite{zmitri2020magnetic, 9403984}.
 
Depending on the specific tasks, the DT models may aim at estimating features to varying degrees of confidence, e.g., the accuracy in estimation of position and velocity at different levels may be depend both on the model goals and the measured signals from sensors. Under this premise, it is fundamental to have efficient policies for selecting appropriate $\SA$s based on their measurement errors, the requirements of the DT model, and the cost of transmitting observations over a wireless network. 
The problem of scheduling IoT sensors based on Value of Information (VoI) under given underlying communication constraints has been considered in \cite{hashemi2020randomized, 9656153}, with the aim to minimize the MSE of the state estimate with imprecise measurements. The authors in \cite{9768131} proposed a way to schedule sensing agents based on VoI to maximize the accuracy of various summary statistics of the state, which has potential applications in industrial automation or safety. 
The aforementioned works and related literature thereof \cite{hashemi2020randomized, 9656153, 9768131}, however, do not consider the incurred communication costs or evaluate the importance of each feature within the state space of $\PA$. Furthermore, measurement errors of sensing agents have not been taken into account, which is critical.

This paper formulates a weighted optimization problem for jointly minimizing errors in state estimation and the total power consumption in a wireless system such as the one in Fig.~\ref{fig_system_model}, as well as defining a scheduling algorithm for $\SA$s under their observations' qualities and communication resource constraints for predictive maintenance of the DT model.
The confidence requirements in estimating each feature are assumed to be heterogeneous: the acceptance margins of estimation errors are individualized and have different priorities. This leads to a different number of scheduled $\SA$s, depending on the VoI of their observations. Our contributions are listed as follows: \textit{(i)}, we propose a DT system tracking dynamic changes of the system parameters and formulate a novel optimization problem to efficiently schedule $\SA$s for maintaining the confidence of system estimate of the DT while minimizing the energy cost; \textit{(ii)}, we propose a VoI-based framework for efficient and practical implementations yielding a low-complexity solution in polynomial time; and \textit{(iii)}, we evaluate the proposed algorithm via numerical simulations, which show that it outperforms the other benchmarks in terms of both computational complexity and power consumption, while providing an improved or equal DT estimation error.

\section{System  Model}
{We consider a WSN, as shown in Fig.~\ref{fig_system_model}, including one $\PA$ and a set of $\Mcal=\{1,2,\dots,M\}$ $\SA$s. The $\SA$s observe the environment and communicate with the \emph{access point} (AP) through a time-slotted wireless channel for building the DT model of $\PA$, where each \emph{query interval} (QI) occurs at $n \in \Ncal= \{1, 2, \dots, N\}$. These $\SA$s synchronize their DTs including locations and power budget status with the Cloud platform. Based on the current model state and the updated information of the environment, the DT model predicts the state of the $\PA$, the optimal policy on scheduling $\SA$s, and the required transmitting power. This feedback signal is sent to the AP to take an action in the physical world.} In the following, upper- and lower-case bold letters denote the matrices and vectors, respectively.  $\mathbf{x}\sim\mathcal{CN}(\boldsymbol{\mu}, \boldsymbol{\Sigma})$ denotes a random vector $\mathbf{x}$ following a complex circularly symmetric Gaussian distribution with mean $\boldsymbol{\mu}$ and covariance matrix $ \boldsymbol{\Sigma}$. $\mathbb{E}[\cdot]$ is the  statistical expectation of the argument.

The $\PA$ operates in a $K$-dimensional process $\Kcal=\{1,2,\dots,K\}$, whose state at QI $n$ is denoted as $\bs(n) = [s_1(n), s_2(n), \dots, s_K(n)]^T$, and evolves as follows:
\begin{align}
	\bs(n) &= f(\bs(n-1)) + \bu(n), \forall n\in\Ncal, 
\end{align}
where $f:\mathbb{R}^{K}\rightarrow\mathbb{R}^K$ is the state update function and $ \bu(n)\sim \mathcal{N}(\mathbf{0},\mathbf{C}_{\bu})$ stands for the process noise. 
At QI $n$, the $\SA$ $m \in \mathcal{M}$ receives a $D$-dimensional observation $\bo_m(n)\in\mathbb{R}^{D}$ of the $\PA$'s state (with $D\leq K$) as
$\bo_m(n)= \bH_m\bs(n)+ \bw_m(n), \forall m\in\Mcal,$ 
where $\bH_m\in\mathbb{R}^{D\times K}$ is  the observation matrix, and $\bw_m(n)\sim \Ncal(\mathbf{0}, \mathbf{C}_{\bw_m})$ stands for the measurement noise. In general, the covariance matrices $\mathbf{C}_{\bu}$ and $ \mathbf{C}_{\bw_m}$ are not diagonal. 
We assume that $\{\{\bs(n)\}_{\forall n},\{\bu(n)\}_{\forall n}, \{\bw_m(n)\}_{\forall m, n}\}$ are uncorrelated, and that $f$, $\mathbf{C}_{\bu}$, and  $\bH_m,\mathbf{C}_{\bw_m}\,\forall m$ are known.
\subsection{Communication System}
\begin{figure}[t!]
	\centering
	\includegraphics[width=0.4\textwidth]{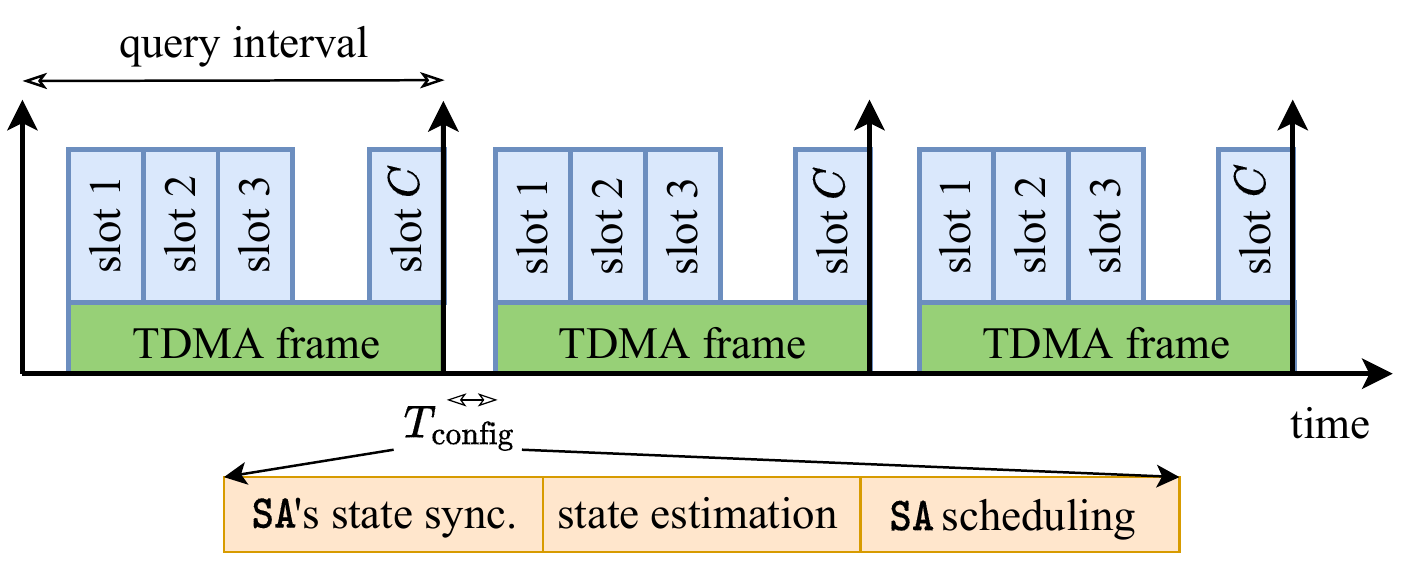}
	\vspace*{-0.3cm}
	\caption{Communication diagram.}
	\label{fig_tdma}
	\vspace*{-0.65cm}
\end{figure}
The communication between the AP and $\SA$s uses a Time Division Multiple Access (TDMA) access scheme, with alternating downlink and uplink phases, and the communication link between AP and cloud is assumed to be perfect. In Fig.~\ref{fig_tdma}, the parameter $T_\text{config}$ accounts for the time the DT model requires to update the active state of $\SA$s, after which it estimates the full state and schedules a maximum of $C$ $\SA$s via the AP in the downlink phase. Then, these scheduled $\SA$s use the uplink phase to forward their observations within the TDMA frame. Consequently, the DT model periodically updates the state of the PA following each query interval. 

We consider block fading channels for the AP$-\SA$ links, in which the channel remains unchanged within each fading block and includes both a line of sight component, experiencing small-scale fading with rich scattering \cite{8698468}. The channel between AP and $\SA$ $m$ is then modeled as
	$h_m(n)=\sqrt{\mu_m} g_m(n), $ 
where the large-scale average channel power $\mu_m$ stands for signal attenuation due to both the path loss and shadowing, and $g_m(n) = \sqrt{\frac{G}{G+1}}\bar{g}+\sqrt{\frac{1}{G+1}}\Tilde{g}$ is the small-scale fading coefficient, and $G$ indicates the Rician factor. For an average channel power gain at the reference distance $\mu_0$ and path loss exponent $\alpha$, the path loss is $\mu_m=\mu_0\hat{d}_m^{-\alpha}$. Thus, given bandwidth $W$, the Signal-to-Noise Ratio (SNR) $\gamma_m(n)$ between the AP and the $m$-th $\SA$ is 
	$\gamma_m(n) = ({p^\text{tx}_m(n)\mu_0|g_m(n)|^2})/({\hat{d}_{m}^{\alpha}\text{W}N_0}),$ 
where $N_0$ represents the noise power spectral density, $p^\text{tx}_m(n)$ is the transmission power, and $\hat{d}_m$ is the distance between the AP and the $m$-th $\SA$. We apply Shannon's bound to obtain the capacity $R_m(n)$ of the link, which is
$R_m(n) = \text{W}\log_2(1+\gamma_m(n) ).$ However, since uplink data are transmitted at a fixed rate $R^{\mathtt{th}}$ due to the simplicity of the sensors, we consider the  reliability condition
	$\mathbb{P}[R_m(n) < R^\mathtt{th}]\leq \varepsilon,$ 
where $\varepsilon>0$ is a system parameter~\cite{9113440}.
\begin{lemma}
	For a given bandwidth $W$ and distance $\hat{d}_m$, the transmitted power of $\SA$ $m$ for reliable transmission with maximum packet erasure probability $\varepsilon$ is upper bounded by:
	\begin{equation}\label{power_require}
		p^\text{tx}_m(n) = \frac{2 WN_0(1+G)(2^{R^\mathtt{th}/W}-1)}{y_Q^2\mu_m},
	\end{equation}
	where $y_Q = \sqrt{2G}+\frac{1}{2Q^{-1}(\varepsilon)}\log(\frac{\sqrt{2G}}{\sqrt{2G}-Q^{-1}(\varepsilon)})-Q^{-1}(\varepsilon)$, and $Q^{-1}(\cdot)$ is the  inverse Q-function.
\end{lemma}
	The proof  is omitted due to space limitations (see \cite{8017572}). 
\subsection{Problem Formulation}
The DT model aims at  maintaining an accurate estimate of $\PA$'s state over its belief of states. 
Herein, the predicted estimator $\hat{\bs}(n)$ of ${\bs}(n)$ is modeled with $p(\bs(n))\sim \mathcal{N}(\hat{\bs}(n), \bPsi(n)), n \in\Ncal$, 
where the covariance matrix $\bPsi(n)$ is updated at QI $n$ according to a Kalman Filter (KF). The MSE of the estimator is
	$\text{MSE}_{} = \mathbb{E}\big[||\bs(n)-\hat{\bs}(n)||^2_2\big], n \in\Ncal$. 
\begin{remark}\label{certainty}
	{\color{black}
		We define the maximum acceptable standard deviation for feature $k\in\Kcal$ as $\xi_k$. This corresponds to the following condition:
		\begin{equation}\label{qos_condition}
			[\bPsi(n)]_{k} \leq {\xi}^2_{k}, \forall k \in\Kcal,
		\end{equation}
		where $[\bPsi(n)]_{k}$ is the $k$-th element of the diagonal of $\bPsi(n)$.}
\end{remark}
Since the $\SA$s communicate over a wireless channel with the AP, the $\PA$ cannot reach all of the $\SA$s at any given time. The reachable subset of $\SA$s at QI $n$ by $\Pcal(n)$ is defined as
\begin{equation}\label{Pcal_compute}
	\mathcal{P}(n)\triangleq\{{{m}}\in\Mcal: d_{m}\leq d_{\max}\},
\end{equation}
with $d_{\max}$ indicating the maximum sensing range of $\SA$s. The scheduling decision is then determined by two factors: the first is the expected VoI of each $\SA$, which affects the accuracy of the $\PA$'s future estimates of the state~\cite{box2011bayesian}, while the second is the quality of the link between it and the $\PA$, impacted by the distance.
We then pose a joint scheduling and power control problem, in which we select the $\SA$s that should transmit and minimize the power, whilst meeting the reliability constraints \eqref{qos_condition}. First, we define the scheduling set $\Qcal(n)$:
\begin{equation}
	\Qcal(n)=\left\{m\in\mathcal{P}(n):p^\text{tx}_m(n)>0\right\}.
\end{equation}
In order to fit the $\SA$s in a QI, we define a maximum number of connections $C$, corresponding to the number of uplink TDMA slots before the next QI. The problem is then defined as follows, using the power allocation vector $\mathbf{p}^\text{tx}(n)=\{p^\text{tx}_m\}$ as the optimization variable:
\begin{subequations} \label{glob_problem}
	\begin{alignat}{2}
		\mathbf{p}^*=&\  \underset{\mathbf{p}^\text{tx}(n)}  { \mathrm{argmin}}& &(1-\alpha) \sum_{k\in\Kcal}\max\left\{\frac{[\bPsi(n)]_{k }}{\xi^2_{k}}-1,0\right\} \nonumber\\
		&&&+\alpha \sum_{\mathclap{m\in\Pcal(n)}}p^\text{tx}_m(n); \label{glob_problema}\\
		&\mbox{such that } && |\Qcal(n)| \leq C, \label{glob_problemc}\\
		&&& \mathbb{P}[R_m(n) < R^\mathtt{th} ]\leq \varepsilon, \forall   {m}\in \Qcal(n), \label{glob_problemd}
	\end{alignat}
\end{subequations}
where the non-negative parameter $\alpha\in[0,1]$ represents the relative weight of the accuracy and energy efficiency in the objective function. It is clear that by querying observations from more $\SA$s, the estimation accuracy increases, at the expense of energy efficiency. For those $\SA$s with significant errors in their measurements, or have features that will not contribute in satisfying the confidence requirements of the $\PA$ (i.e., the ones with a low VoI), measuring and sending observations consumes unnecessary energy, which should taken into account. 
The constraints in \eqref{glob_problemc},~\eqref{glob_problemd} are required to ensure that transmissions can be reliably performed within a TDMA uplink slot, that no more than $C$ $\SA$s transmit in any QI.
We emphasize that problem \eqref{glob_problem} is a non-convex program due to the non-convexity of the objective function \eqref{glob_problema} and of constraints  \eqref{glob_problemc},~\eqref{glob_problemd}. Furthermore, the selection of nodes makes the problem equivalent to the classic knapsack problem, making finding the solution NP-hard. Accordingly, a heuristic algorithm is used to obtain an efficient suboptimal solution.

\section{VoI Sensing Agent Selection}

The MMSE estimator for a KF is given in \cite[Eq. (1)]{Kalman}, and since our objective is to minimize the (weighted) variance of components of the state, we use the standard Kalman estimator. It is worth noting that, since the general system is nonlinear, the Kalman equations are a linearization of the true dynamic system, and as such, might lead to additional errors. The use of the Extended Kalman Filter (EKF) is common in the IoT literature~\cite{huang2019epkf}. Another important assumption is that the virtual world has full knowledge about the statistics of the process, i.e., the update function $f(\mathbf{s})$ and the noise covariance matrices. This is also commonly assumed in the relevant literature, as the system statistics estimation can be performed before deployment. As joint power allocation and VoI maximization is excessively complex, we use the heuristic of splitting into separate scheduling and power allocation problems. The main idea to solve \eqref{glob_problem} is that at each QI $n$, the minimum number of $\SA$s is selected to transmit to maintain the estimation certainty of state $\bs(n)$ at the required level.

The heuristic steps are as follows: first, we determine the set of $\SA$s $\mathcal{Q}^*(n)$ with minimal cardinality which satisfies constraint~\eqref{glob_problemc} and for which all selected features satisfy the condition in~\eqref{qos_condition}; the minimum power to establish a reliable communication link between scheduled $\SA$s and  AP is then set as in~\eqref{power_require}.
The heuristic, whose pseudocode is listed in Algorithm \ref{alg_global},  effectively addresses problem \eqref{glob_problem}.

In the dynamic scenario, the initial state $\bs(n)$ is a random vector with certain mean $\mathbb{E}[\bs(n)] = \boldsymbol{\mu}_{\bs(0)}$ and covariance matrix $\text{Cov}[\bs_0]= \mathbf{C}_{\bs(0)}$. $\Qcal(n)$ is initialized as an empty set due to no prior information. The EKF then computes the estimation errors for the belief  $\hat{\bs}(n) \sim \mathcal{CN}(\bmu_{\hat{\bs}(n)},\bPsi^{\mathtt{pr}}(n))$  at the $\PA$  based on prior updates $\hat{\bs}(n-1)$ as
\begin{align}\label{prior_error}
	\bPsi^{\mathtt{pr}}_{\hat{\bs}}(n) = \bP\bPsi_{\hat{\bs}}(n-1)\bP^T + \bC_{\bu(n)}.
\end{align}
For given error variance qualities $\bepsilon \triangleq \{\xi_{k}\}_{k\in\Kcal}$, the conditions in \eqref{qos_condition} result in two possible cases: (1) If those conditions hold for all $k\in\Kcal$, the DT model satisfies the required bound without receiving any observation from the $\SA$s. The prior update is sufficient to ensure the confidence in estimate and $\Qcal^*(n) = \emptyset$; (2) 
If one of those conditions is violated, at least one interesting feature is not estimated accurately enough, and we need the corresponding observations to improve the estimation, as scheduled by our heuristic. We stress that in the first case, the belief can be computed using the EKF blind update operation:
\begin{equation}\label{mu_prior}
	{\hat{\bs}^{\mathtt{pr}}(n)}  = \bP{\hat{\bs}(n-1)} + \bmu_{\bu}.
\end{equation}
\begin{algorithm}[t]
	\begin{algorithmic}[1]{\fontsize{8pt}{9pt}\selectfont
			\protect\caption{$\SA$ scheduling algorithm for problem  \eqref{glob_problem}} 
		\label{alg_global}
		\global\long\def\algorithmicrequire{\textbf{Input:}}
		\REQUIRE $\bb_{0,\bo_0}, \bmu_{\bu_0}, C_{\bu_0}$
		Available uplink slots $C$, Power budget $\bar{P}^{\text{tx}}$, The state and requirement certainty $\big(\bs, \{\xi_{k}^2\}\big)$
		\global\long\def\algorithmicrequire{\textbf{Output:}}
		\REQUIRE The scheduled user set $\{\Qcal^*(n)\}$; their belief $\{\hat{\bs}^*(n), \bPsi^*(n)\}$, and the associated transmit power $\{p^\text{tx}_m(n)\}$
		\STATE Initial $\mathcal{Q}(n) = \emptyset$
		\STATE Compute the prior errors $\bPsi^{\mathtt{pr}}(n)$ as in \eqref{prior_error} 
		\IF {$[\bPsi^{\mathtt{pr}}(n)]_{k} \leq \xi_{k}^2, \forall k$}
		\STATE Compute ${\hat{\bs}^{\mathtt{pr}}(n)} $ as in \eqref{mu_prior}
		\STATE Update ${\hat{\bs}(n)}  = {\hat{\bs}^{\mathtt{pr}}(n)} $ and $\bPsi_{\hat{\bs}}(n) = \bPsi_{\hat{\bs}}^\mathtt{pr}(n)$
		\ELSE
		\STATE Set $t=1$ and compute available $\SA$ set $\Pcal(n)$ by  \eqref{Pcal_compute}
		\WHILE{conditions \eqref{checking3} hold}
		\STATE Update $\Qcal(n)$ and $\Pcal(n)$ as in \eqref{update_set}
		\STATE Update the $\bK(n)$, $\bH(n)$ and $C_{\bw(n)}$ as in \eqref{H_compute}, and \eqref{Cw_compute}
		\STATE  Set $t=t+1$
		\ENDWHILE
		\STATE Update $\Qcal^*(n) = \Qcal(n)$
		\STATE Compute ${\hat{\bs}(n)}  = {\hat{\bs}^{\mathtt{pos}}(n)} $ and $\bPsi_{\hat{\bs}}(n) = \bPsi_{\hat{\bs}}^\mathtt{pos}(n)$ as in\eqref{pos_update},   \eqref{pos_variance}
		\STATE Compute the power consumption $\{p^\text{tx}_m(n\}$ as in \eqref{power_require}
		\ENDIF
	}
\end{algorithmic}
\end{algorithm}
In the second case, we run Algorithm 1. 
It is worth noting that  at the $t$-th iteration, if any constraint is still unmet and $\{|\Qcal(n)| <\ C, |\mathcal{P}(n)| >0\}$, there is room for scheduling new $\SA$s to join $\Qcal(n)$. In order to be able to choose the most uncertain candidate feature $s_k^{*}(n), k\in\Kcal$, the following optimization problem is considered at the $t$-th iteration: 
\begin{subequations} \label{finding_state}
\begin{alignat}{2}
	s_k^{*}(n) =& \underset{s_k(n)\in\bs(n)}{\ \mathrm{argmax }} &\quad & [\bPsi^{(t)}(n)]_{k }/ \xi^2_{k} \\
	&\mbox{subject to} &&  {m} \in\mathcal{P}(n), \forall m\in\Mcal, \label{finding_stateb}\\
	&&& {m} \rightarrow s_{k}(n), \forall m\in\Mcal,
\end{alignat}
\end{subequations}
where ${m} \rightarrow s_{k}(n)$ means that the $\SA$ $m$ measures feature $s_k(n)$.
At iteration $t=1$, we set $\bPsi^{(1)}_\bs(n) = \bPsi^{\mathtt{pr}}_{\hat{\bs}}(n)$. We emphasize that according to constraint \eqref{finding_stateb}, feature $s_k^{*}(n)$ is selected only if at least one $\SA$ ${{m}}\in \Pcal(n)$ can provide coordinating observations. Then, the  $\SA$ ${{m}}^*\subset\mathcal{P}(n)$ measuring feature $s_k^{*}(n)$ with the minimum error covariance is chosen to send its measurement. The scheduled and available $\SA$ sets $\Qcal(n)$ and $\mathcal{P}(n)$ are updated as
\begin{equation}\label{update_set}
\Qcal(n) \leftarrow \Qcal(n) \cup\{ {{m}}^* \};\  \Pcal(n) \leftarrow\Pcal(n)\backslash\{{{m}}^*\}.
\end{equation}
 $\bH(n)$ and $\bC_{\bw(n)}$ are the combination observation and covariance matrices, which are respectively formulated as
\begin{align}
\bH(n) &= [\bH_{1};\bH_2;\dots;\bH_{|\Qcal(n)|}], \label{H_compute} \\
\bC_{\bw(n)} &=  \text{diag}[\bC_{\bw_1}, \bC_{\bw_2},\dots,\bC_{\bw_{|\Qcal(n)|}}]\label{Cw_compute},
\end{align}
where $\bH_{{m}}$ is the  observation matrix of  the $\SA$ ${{m}} ({{m}}\in\Qcal(n))$.
The posterior error covariance matrix is derived by
\begin{align}\label{pos_variance}
\bPsi^{\mathtt{pos}}_{\hat{\bs}}(n)=  (\bI - \bK(n)\bH(n))\bPsi^{}_{\hat{\bs}}(n-1),
\end{align}
where $\bK(n)$ is the EKF gain, computed using the standard KF equation.
The iterative loop continues as long as all three of the following conditions are true: 
\begin{align}
\{|\Qcal^*(n)| &< C;\  
\exists [\bPsi(n)]_{k} \geq {\xi}^2_{k}; \ 
\exists s_k^{*}(n) \mbox{ in }\eqref{finding_state}\}. \label{checking3}
\end{align}
{\color{black}Hence, it makes intuitive see that the loop repeats for at most $C$ iterations before terminating.} The posterior update is then:
\begin{equation}\label{pos_update}
{\hat{\bs}^{\mathtt{pos}}(n)}  = {\hat{\bs}^{\mathtt{pr}}(n)} + \bK(n)(\bo(n) - \bH(n){\hat{\bs}^{\mathtt{pr}}(n)} ),
\end{equation}
where $\bo(n) $ represents the combination of received $\SA$ observations. Accordingly, we update $\hat{\bs}(n) = \hat{\bs}^{\mathtt{pos}}(n)$. Despite the local $\SA$ scheduling solution, our  approach ensures the long-term balance between state certainty and communication cost over different QIs with respect to the $\PA$'s requirements. 
\begin{figure*}[t]
\begin{minipage}{0.9\textwidth}
	\centering
	\includegraphics[trim=0.cm 15.5cm 8.5cm 0.8cm, clip=true, width=4.1in]{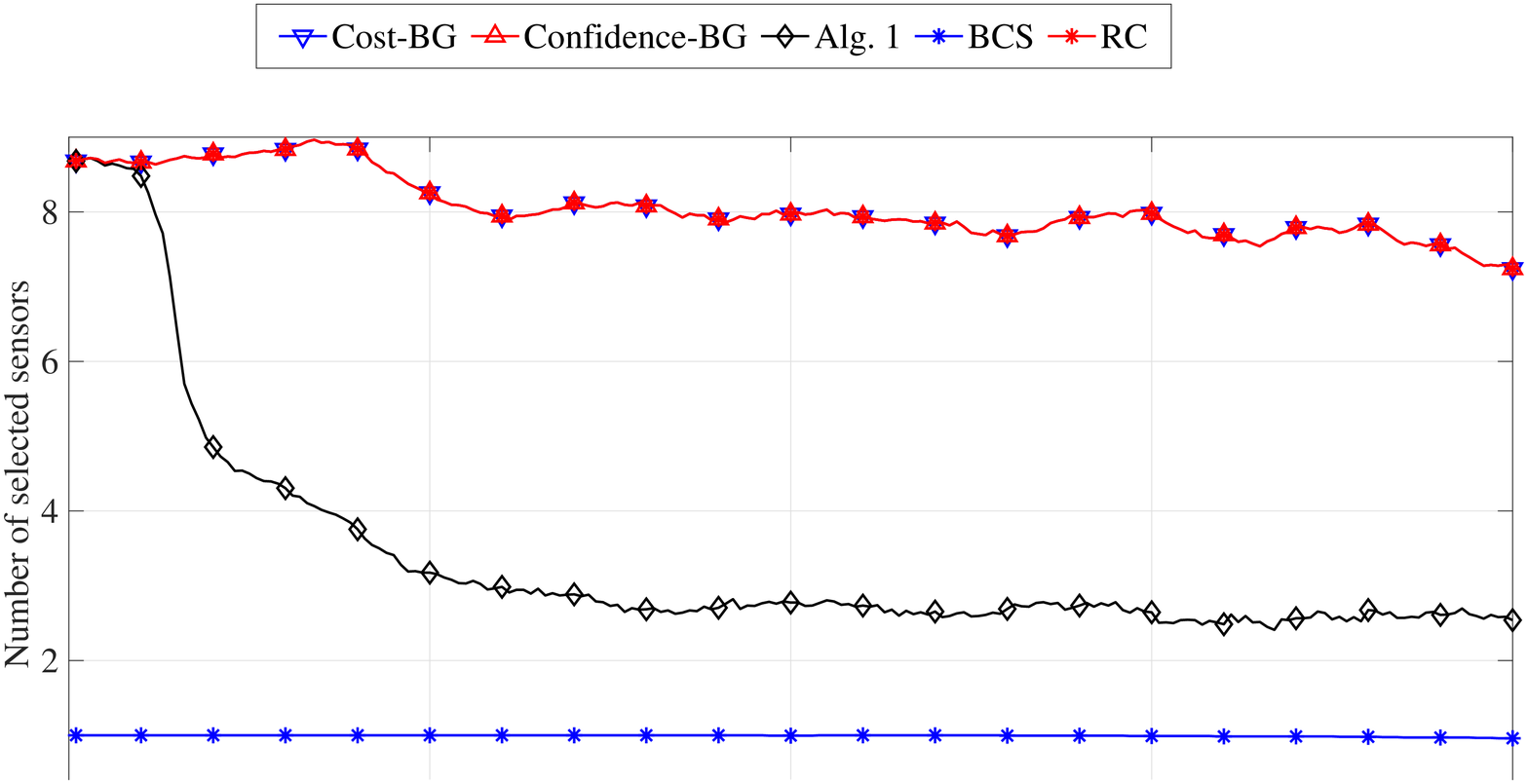} \\ 
	\vspace*{-0pt}
\end{minipage}
\begin{minipage}{0.24\textwidth}
	\includegraphics[trim=0.cm 0.1cm 0.cm 0.8cm, clip=true, width=1.3in]{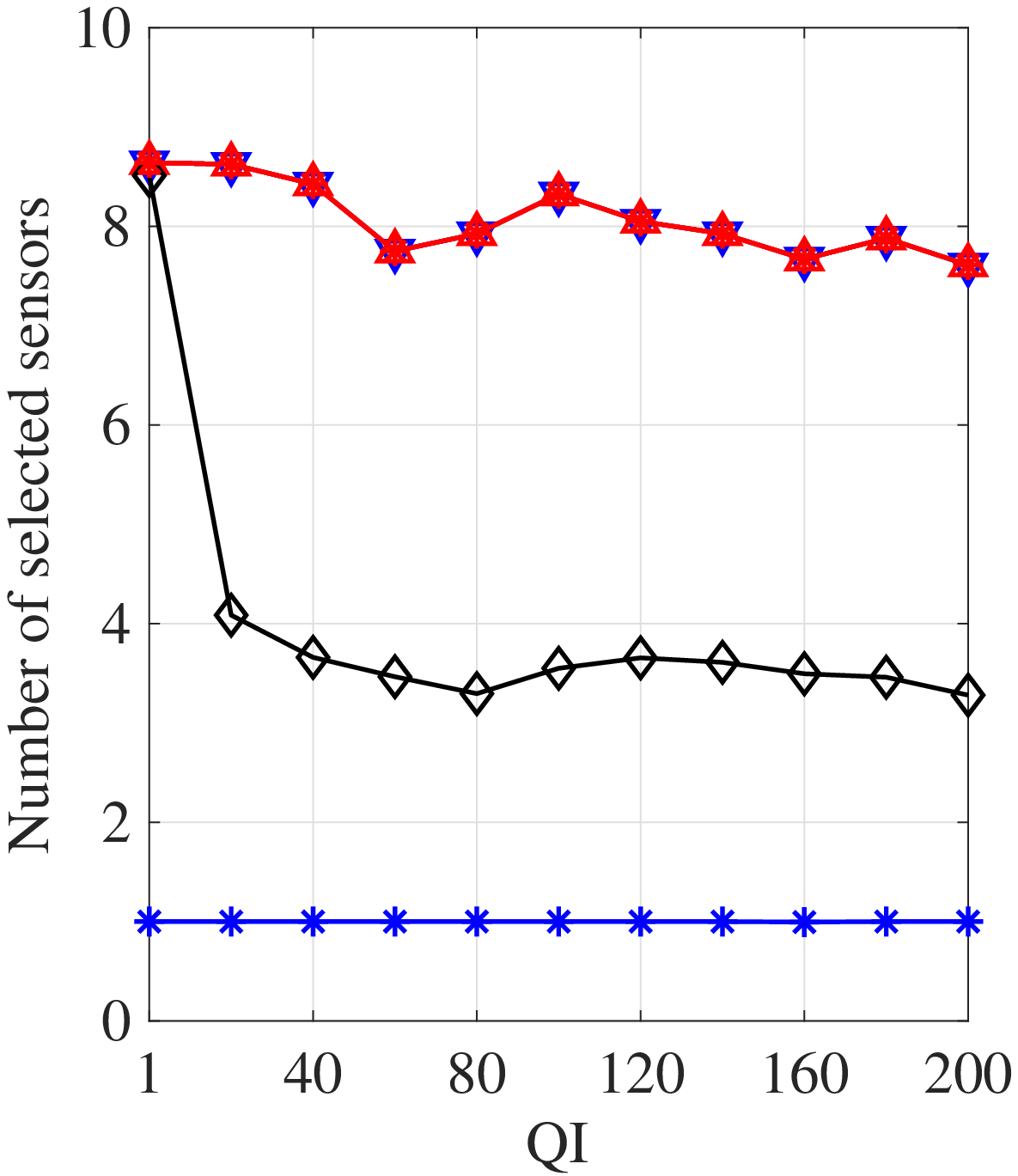} \\ 
	\vspace*{-0pt}
	\centering \fontsize{8}{8}{$(a)$}
	\vspace*{-5pt}
\end{minipage}
\begin{minipage}{0.24\textwidth}
	\includegraphics[trim=0.cm 0.1cm 0.cm 0.8cm, clip=true, width=1.3in]{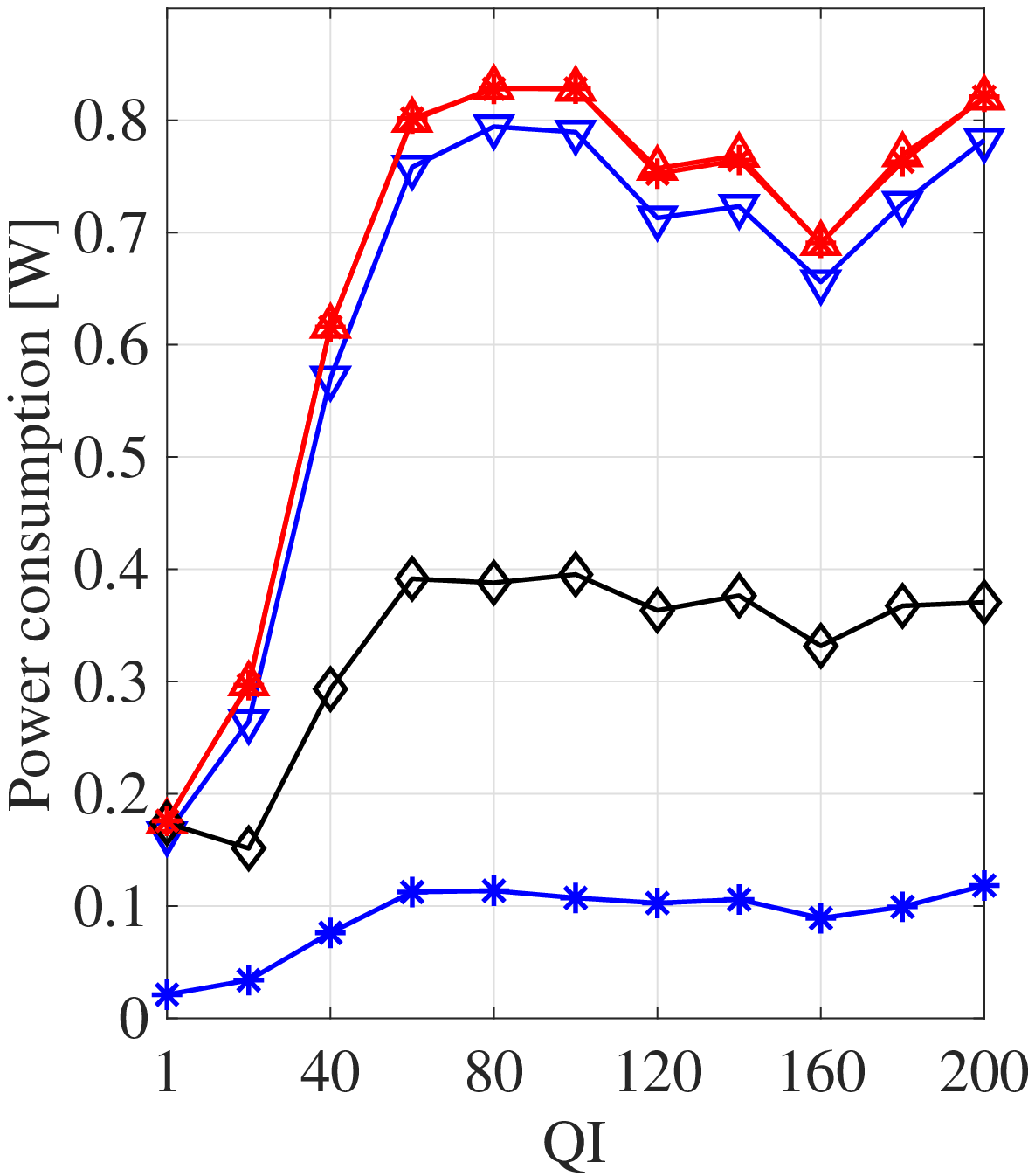} \\ 
	\vspace*{-0pt}
	\centering \fontsize{8}{8}{$(b)$}
	\vspace*{-5pt}
\end{minipage}
\begin{minipage}{0.24\textwidth}
	\includegraphics[trim=0.cm 0.1cm 0.cm 0.8cm, clip=true, width=1.3in]{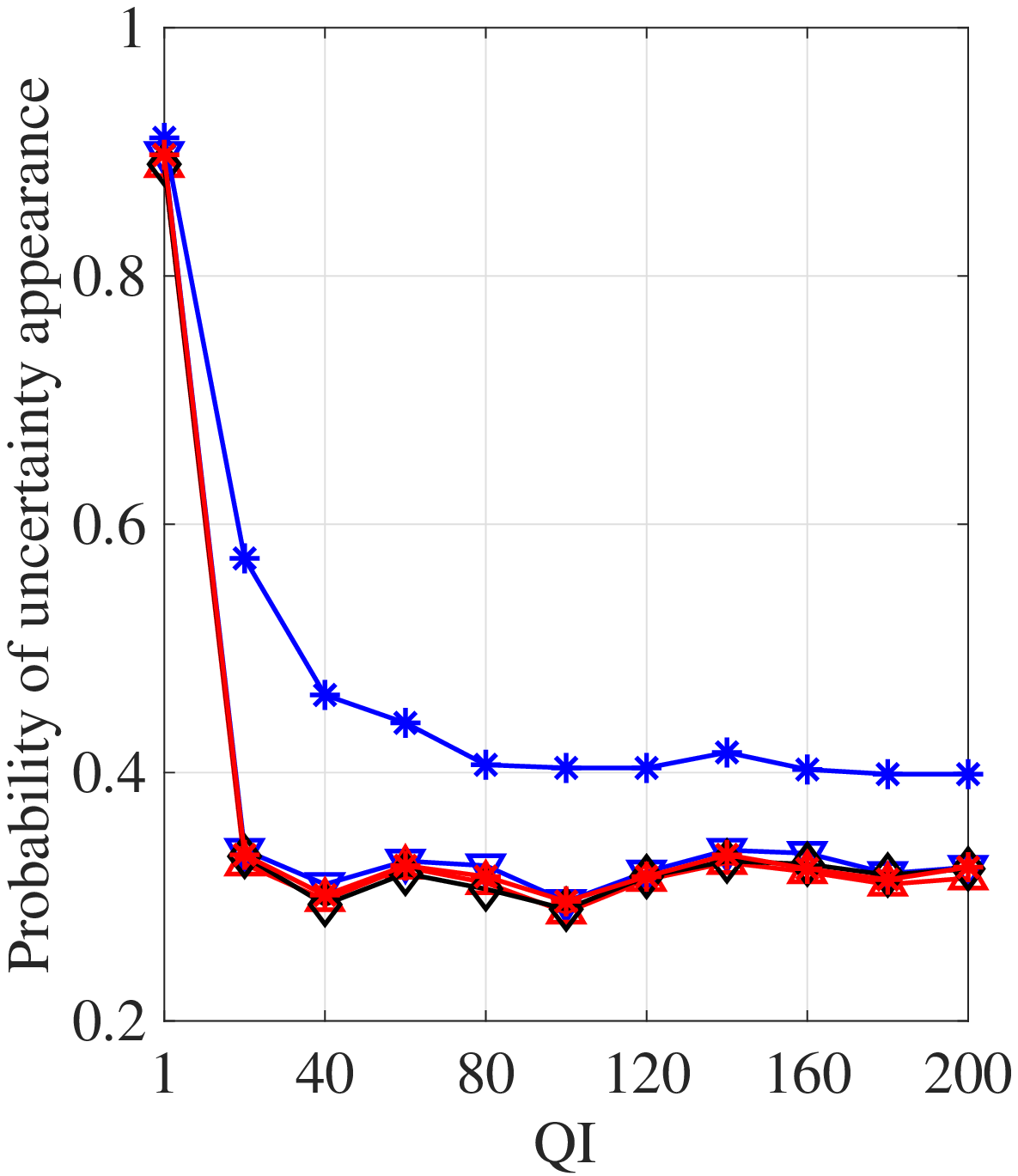} \\ 
	\vspace*{-0pt}
	\centering \fontsize{8}{8}{$(c)$}
	\vspace*{-5pt}
\end{minipage}
\begin{minipage}{0.24\textwidth}
	\includegraphics[trim=0.cm 0.1cm 0.cm 0.8cm, clip=true, width=1.3in]{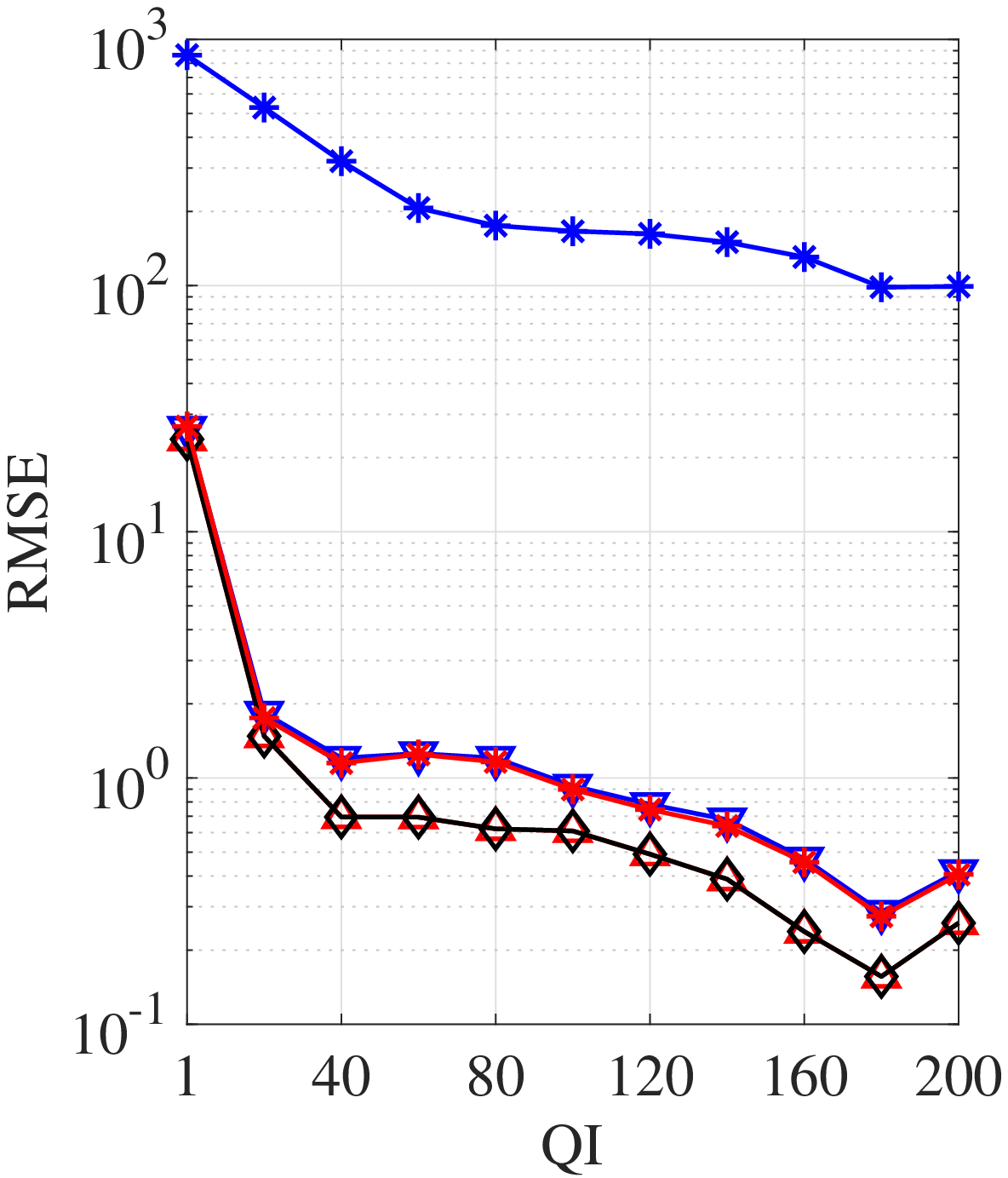} \\ 
	\vspace*{-0pt}
	\centering \fontsize{8}{8}{$(d)$}
	\vspace*{-5pt}
\end{minipage}
\vspace*{-0.05cm}
\caption{The system performance of different metrics versus query interval  ($M = 60$ sensors including $30$ position sensors and $30$ velocity sensors): $(a)$ the number of selected sensors; $(b)$  the power transmission; $(c)$ the probability of  uncertainty appearance; and $(d)$ the RMSE.}
\label{fig:per_vs_QI}
\vspace*{-12pt}
\end{figure*}
\begin{figure*}[t]
\begin{minipage}{0.9\textwidth}
	\centering
	\includegraphics[trim=0.cm 15.5cm 8.5cm 0.8cm, clip=true, width=4.1in]{legend.eps} \\ 
	\vspace*{0pt}
\end{minipage}
\begin{minipage}{0.245\textwidth}
	\includegraphics[trim=0.cm 0.1cm 0.cm 0.7cm, clip=true, width=1.3in]{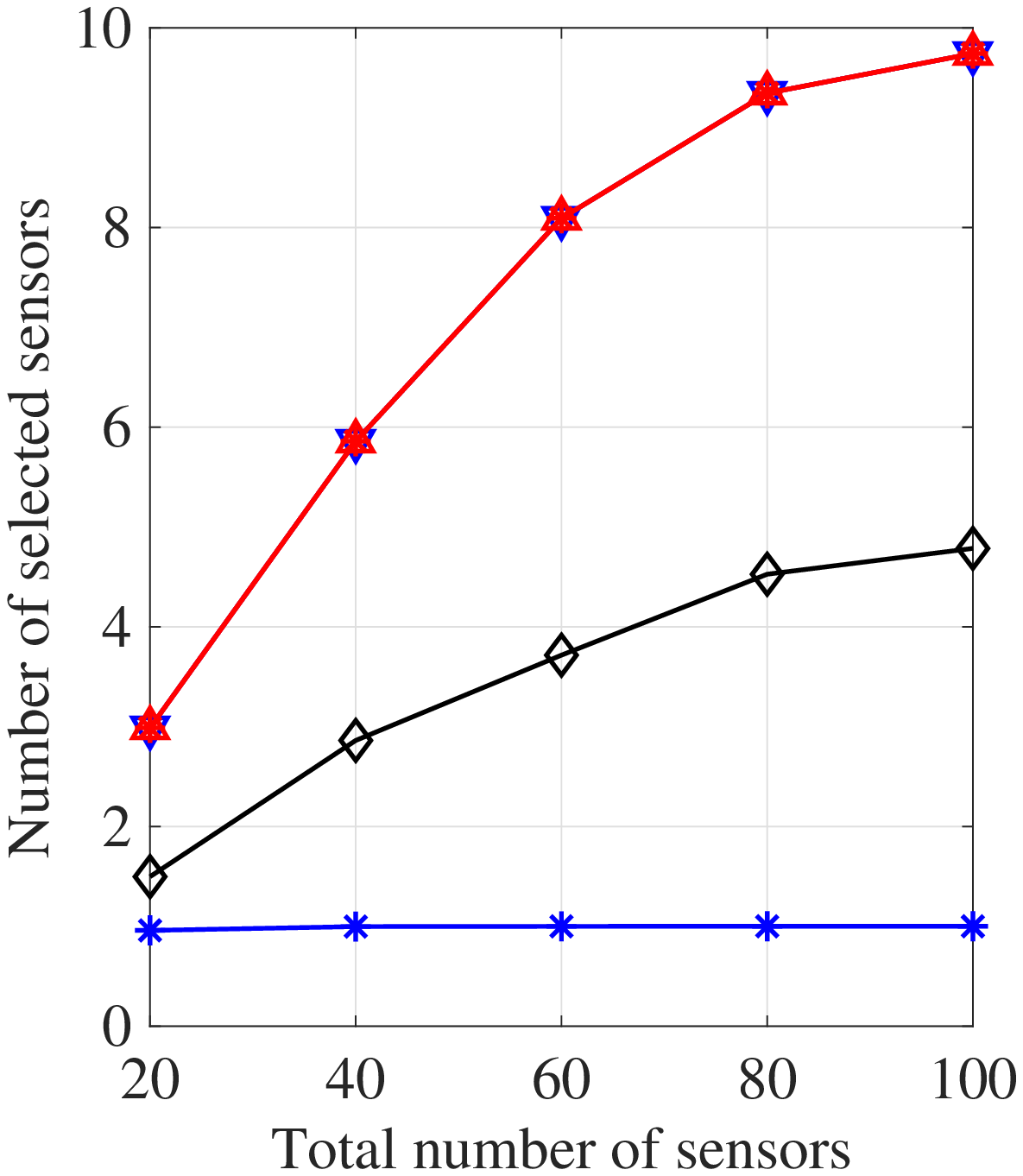} \\ 
	\vspace*{-0pt}
	\centering \fontsize{8}{8}{$(a)$}
	\vspace*{-5pt}
\end{minipage}
\begin{minipage}{0.245\textwidth}
	\includegraphics[trim=0.cm 0.1cm 0.cm 0.7cm, clip=true, width=1.3in]{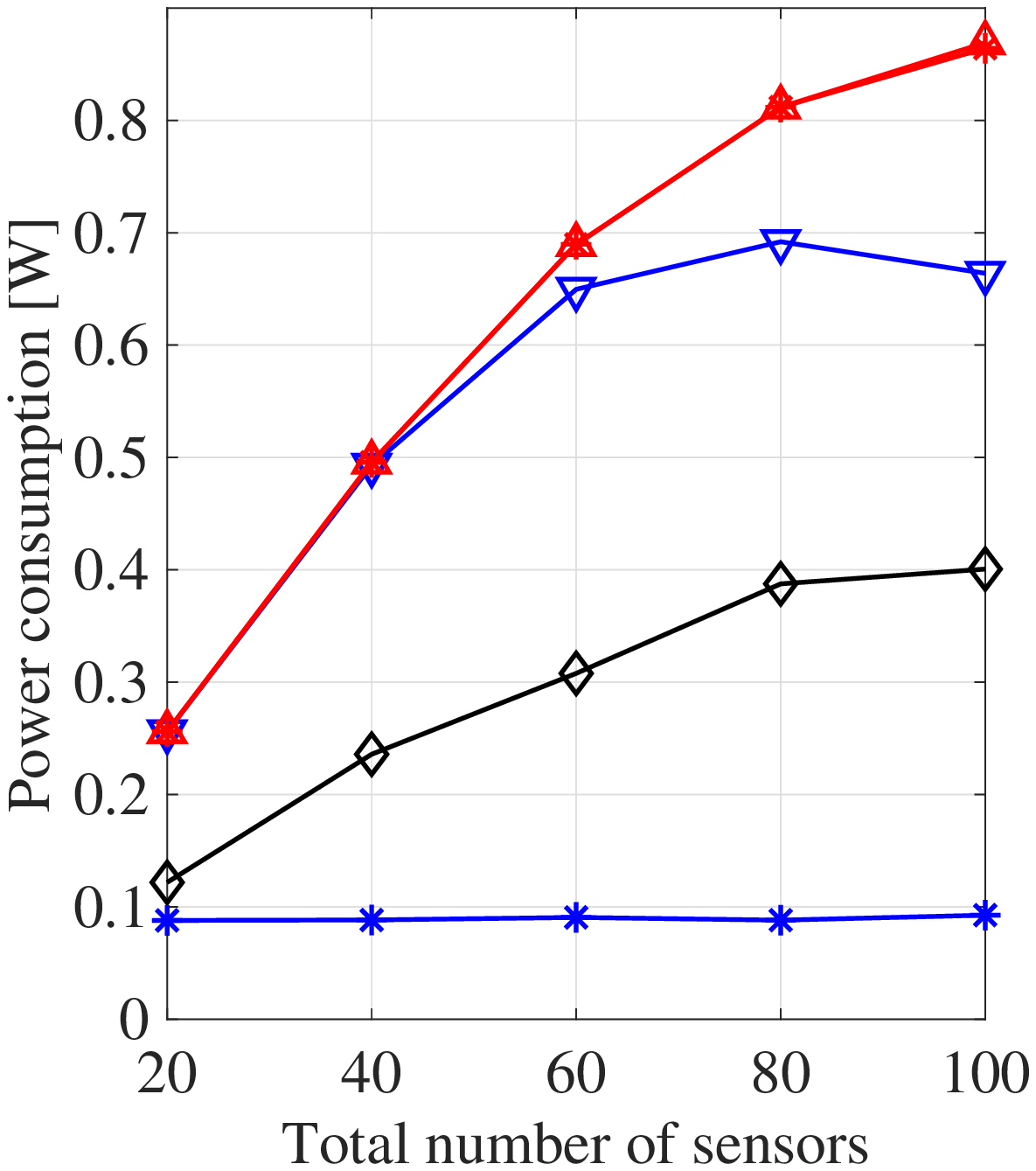} \\ 
	\vspace*{-0pt}
	\centering \fontsize{8}{8}{$(b)$}
	\vspace*{-5pt}
\end{minipage}
\begin{minipage}{0.245\textwidth}
	\includegraphics[trim=0.cm 0.1cm 0.cm 0.7cm, clip=true, width=1.3in]{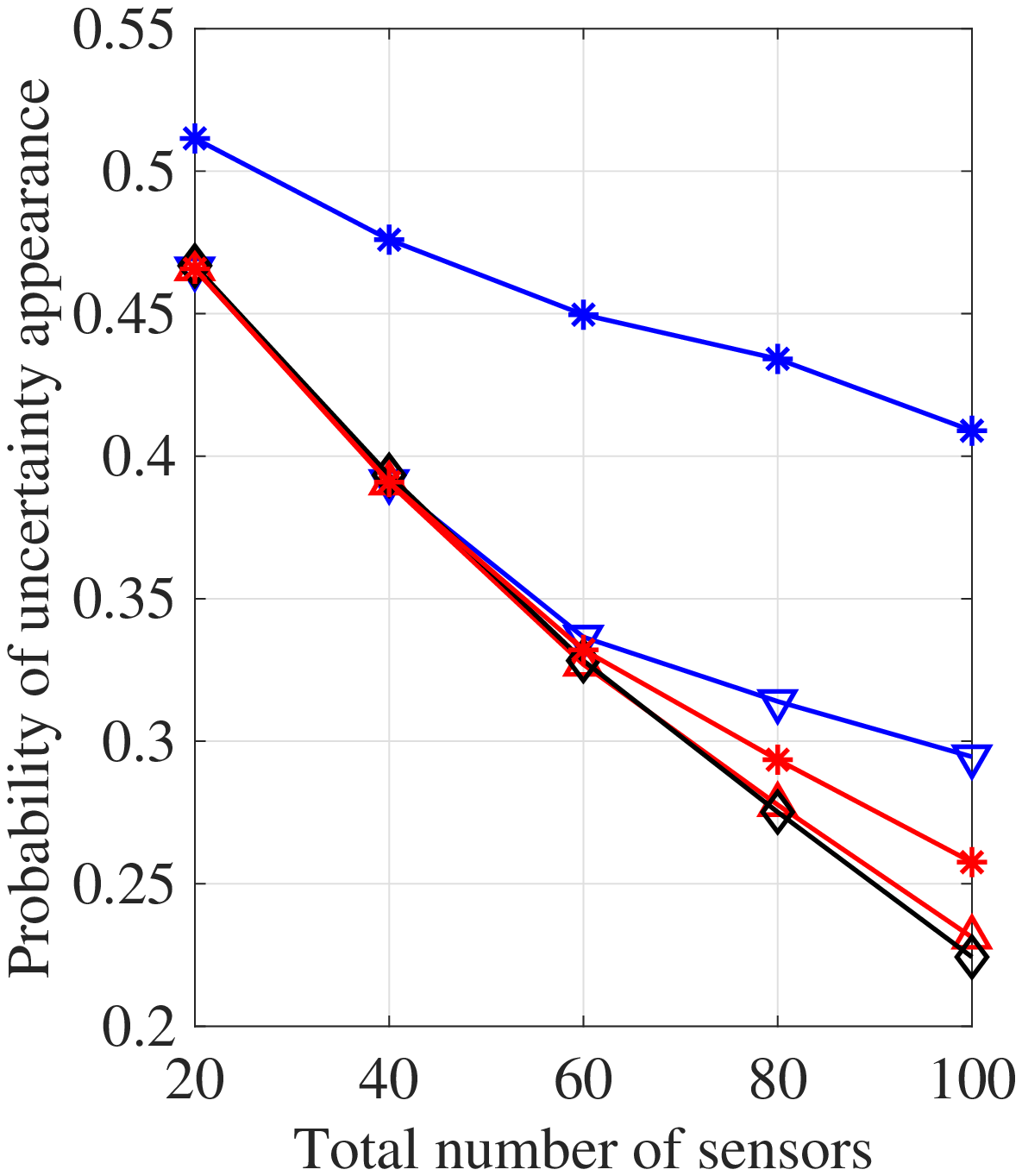} \\ 
	\vspace*{-0pt}
	\centering \fontsize{8}{8}{$(c)$}
	\vspace*{-5pt}
\end{minipage}
\begin{minipage}{0.245\textwidth}
	\includegraphics[trim=0.cm 0.0cm 0.cm 0.7cm, clip=true, width=1.3in]{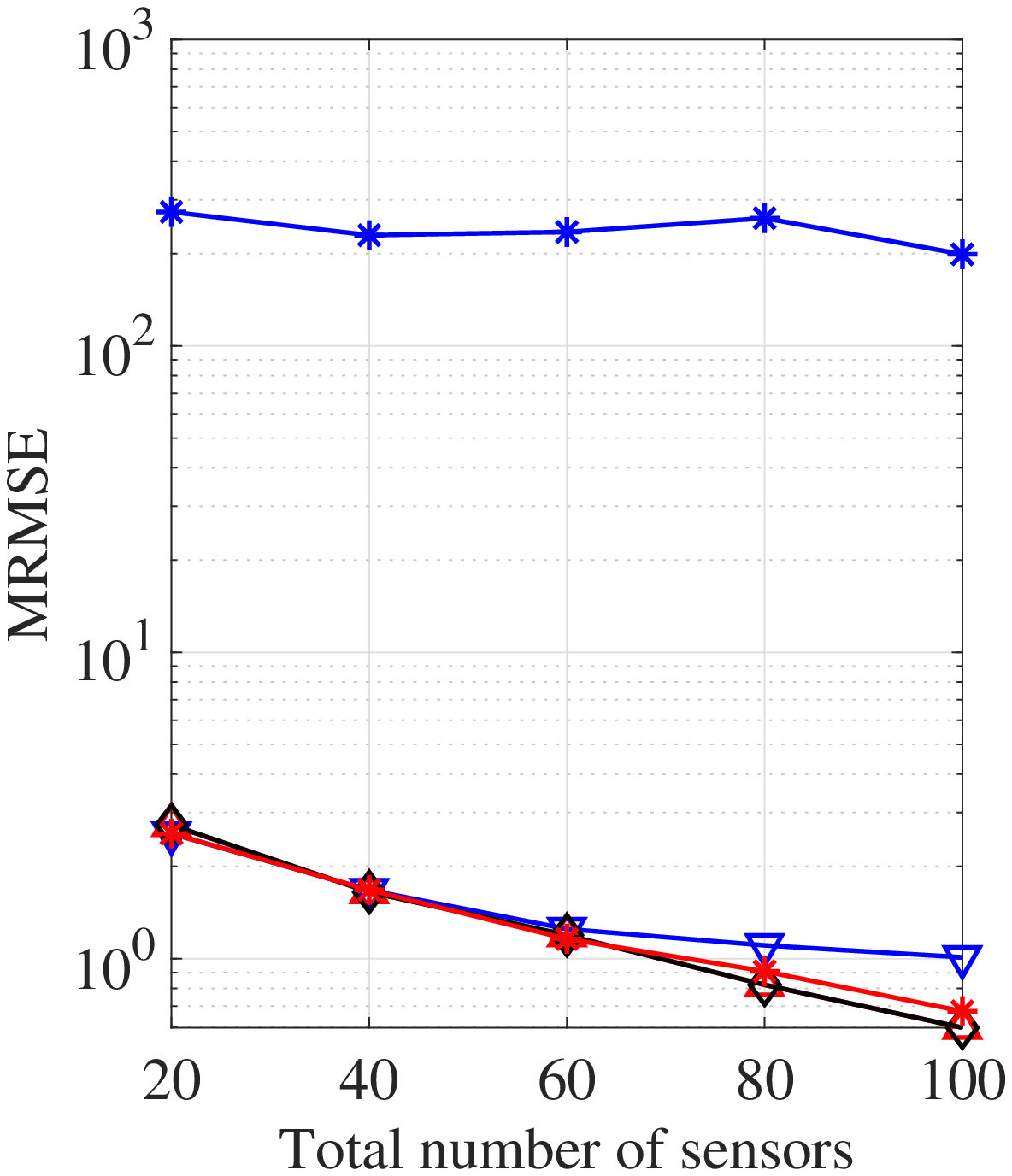} \\ 
	\vspace*{-0pt}
	\centering \fontsize{8}{8}{$(d)$}
	\vspace*{-5pt}
\end{minipage}
\vspace*{-0.2cm}
\caption{System performance of different metrics versus the total number of sensors (equal number of different types of sensors): $(a)$ the number of connections; $(b)$  the power transmission; $(c)$ probability of  uncertainty appearance; and $(d)$ the MRMSE.}
\label{fig:per_vs_num_sen}
\vspace*{-18pt}
\end{figure*}
\section{Numerical Results}
\begin{table}[!t]\vspace{-0.5cm}
\caption{Simulation Parameters}
\resizebox{8.9cm}{!} 
{
	\begin{tabular}{ll|ll}
		\hline
		Parameter & Value & Parameter & Value \\
		\hline
		Carrier frequency ($f_c$)          & 2.4 GHz      &  Rate threshold ($R^\mathtt{th}$) \cite{Neri14}          &   250 kbps    \\
		Bandwidth   (W)       & 5 MHz      &  Channel noise power          & -11.5 dBm     \\
		Required error variances $(\xi^2_{pos}, \xi^2_{vel})$ & (0.015, 0.005)     &  Max. distance ($d_{\max}$)          & 20 m     \\
		Max. connection ($C$) & 10 & $\PA$'s weight ($m_\PA$)& 100 kg\\
		Rician factor ($G$) & 15 dB &  Outage probability factor $(\varepsilon)$& 0.0001\\
		\hline\vspace{-0.5cm}
	\end{tabular}
}
\label{parameters}
\end{table}
In this section, we examine our proposed algorithm in a specific sensor scheduling scenario. Particularly, we consider an object $\PA$ with mass $m_\PA$ moving in the XY circle plane with position vector $\bs(n) = [x(n), y(n)]^T: \mathbb{R}\rightarrow\mathbb{R}^2$, and velocity vector $\bv(n) = [v_x(n), v_y(n)]^T: \mathbb{R}\rightarrow\mathbb{R}^2$, as illustrated in Fig. \ref{fig_system_model}.  To randomly model a movement of $\PA$, we apply a driving force by $\mathbf{f}(n) = [F_x(n), F_y(n)]^T =  [A_x\cos(2\pi f_xn), A_y\cos(2\pi f_yn)]^T$. 
We apply a calibrated additional force, derived experimentally, $\mathbf{g}(n)=\bar{G}\frac{\bx(n-1)-\mathbf{O}}{d_{\PA,\mathbf{O}}}\frac{|\bv(n-1)|}{R-d_{\PA,\mathbf{O}}}$ to the $\PA$, aiming at pulling the $\PA$ into the center whenever the $\PA$ moves close to the edge of the region. Herein, $\bar{G}$ and $\mathbf{O}$ are the force parameter and the center position, respectively. Then, the position, velocity, and acceleration of the $\PA$ are given as $\bx(n)\triangleq\bx(nT) =  \bx(n-1) + T\bv(n-1),  + \frac{n^2}{2}\ba(n-1) + \mathbf{n}_x$; $
\bv(n)\triangleq\bv(nT) = \bv(n-1) + T\ba(n-1) + \mathbf{n}_v$; $
\ba(n)\triangleq\ba(nT) = \frac{\mathbf{f}(n)+\mathbf{g}(n)}{m_\PA},$
where perturbations $\mathbf{n}_x, \mathbf{n}_v$ are modeled as independent random variables, i.e., $\mathbf{n}_x\sim\mathcal{CN}(0,\boldsymbol{\sigma}_{pos}^2)$ and $\mathbf{n}_v\sim\mathcal{CN}(0,\boldsymbol{\sigma}_{vel}^2)$ with $\boldsymbol{\sigma}_{pos}^2=\mathbf{I}\times0.04$ and $\boldsymbol{\sigma}_{vel}^2=\mathbf{I}\times 0.01$. 

The state vector of the $\PA$ includes its position and velocity:
$\bs(n) = [\bx_x(nT)\ \bx_y(nT)\ \bv_x(nT)\ \bv_y(nT)]^T$. 
Other important parameters are included in Table \ref{parameters}.  The observation matrix of position and velocity $\SA$s are respectively given by
\begin{equation}\label{}
\bH_{pos} = \left [ \begin{matrix}
	1 & 0 & 0 & 0\\ 
	0 & 1 & 0 & 0
\end{matrix} \right ]; \quad 	\bH_{vel} = \left [ \begin{matrix}
	0 & 0 & 1 & 0\\ 
	0 & 0 & 0 & 1
\end{matrix} \right ].
\end{equation}
For benchmarking, we compare our proposed Alg. 1 and four other methods: \textit{Cost-based greedy} (Cost-BG) allows the AP to get observations from all nearest $\SA$s based on ascending order of distance from AP to $\SA$ at each QI; \textit{Confidence-based greedy} (Confidence-BG) is similar to Cost-BG but based on descending order of $\SA$s' confidence. The number of scheduled $\SA$s in the greedy benchmarks is $|\Qcal(n)^*| = \min(C, |\Pcal(n)|)$. We also consider \textit{Random scheduling} (RC) , which randomly selects $C$ $\SA$s. Finally, we consider the \textit{Best candidate selection} (BCS) heuristic~\cite{GUPTA2006251}, in which the DT model chooses the best $\SA$ with the highest confidence in $\Pcal(n)$ at each QI.

In Fig. \ref{fig:per_vs_QI}(a) and Fig. \ref{fig:per_vs_QI}(b), we compare  the average number of selected sensors and the  associated power consumption at each QI using different approaches. The results show that Alg. 1 brings benefits to the number of sensors connected and thus saves the used power. At first QIs, Alg 1 utilizes multiple sensors to enhance the efficiency of state estimation. However, after achieving the required estimated confidence, the AP only queries observation from 4 sensors per QI (from QI 30) to maintain the confidence of the estimated state, thereby directly saving power consumption (as in Fig. \ref{fig:per_vs_QI}(b)) compared to other schemes. Fig \ref{fig:per_vs_QI}(c) plots the probability of uncertainty appearance, i.e., the probability of violating at least one of the requirements in \eqref{qos_condition} in the coherence interval. While significantly reducing communication costs, Alg. 1 provides a probability of uncertainty appearance comparable to that of the best greedy algorithm (Confidence-BG). Due to the fact that only one sensor is connected to each QI, the BCS scheme always causes the highest level of uncertainty, even with minimal power consumption, as shownin Fig. \ref{fig:per_vs_QI}(b). By maintaining 4 connections to sensors from QI 40 onward, Alg. 1 can meet the estimation error requirements for all features of object's state, which is approximately $2\times$ lower comparing to Confidence-BG, Cost-BG or RC schemes. The Root Mean Square Error (RMSE) is used in Fig. \ref{fig:per_vs_QI}(c)  \cite{9845870} to evaluate the overall system performance of our algorithm over all benchmarks. It is seen that Alg. 1 always provides the lowest RMSE equivalent to Confidence-BG regardless of QI.

In Fig. \ref{fig:per_vs_num_sen}, we examine the same metrics in Fig. \ref{fig:per_vs_QI} based on the divergence in the number of sensors distributed in the observed area. The average number of sensors connected to the $\PA$ per QI and the corresponding power consumption are shown in Fig. \ref{fig:per_vs_num_sen}(a) and Fig. \ref{fig:per_vs_num_sen}(b), respectively. As expected, Alg. 1 significantly outperform the greedy alternatives in terms of probability of uncertainty appearance. Furthermore, Fig. \ref{fig:per_vs_num_sen}(d) plots the mean RMSE (MRMSE) defined at \cite{9656153} to average the RMSE over the entire simulation time. As the number of sensors allocated increases, our proposed scheme always yields an MRMSE equivalent to the best one (Confidence-BG) and increases the margin compared to others.

\section{Conclusions}
In this paper, we have investigated the optimization of $\SA$s scheduling based on the VoI of sensing agents with limited communication resource. We proposed an efficient algorithm selecting subsets of $\SA$s to meet the requirements in the confidence of state estimation. Future work on the subject may aim at considering the long-term effects of scheduling choices in more complex systems, as well as integrating deep learning-based estimators.
\setstretch{0.91}
\bibliographystyle{IEEEtran}
\bibliography{Journal}
\end{document}